\documentclass[preprint]{revtex4}
\input{epsf}
\begin{document}
\title{Phase slip in a superfluid Fermi gas near a Feshbach resonance}
\author{Lan Yin}
\email{yinlan@pku.edu.cn} \affiliation{School of Physics, Peking
University, Beijing 100871, P. R. China}
\author{Ping Ao}
\affiliation{Departments of Mechanical Engineering and Physics,
University of Washington, Seattle, WA 98195, USA}
\date{\today}
\date{\today}
\begin{abstract}
In this paper, we study the properties of a phase slip in a
superfluid Fermi gas near a Feshbach resonance.  The phase slip
can be generated by the phase imprinting method. Below the
superfluid transition temperature, it appears as a dip in the
density profile, and becomes more pronounced when the temperature
is lowered. Therefore the phase slip can provide a direct evidence
of the superfluid state.  The condensation energy of the
superfluid state can be extracted from the density profile of the
phase slip, due to the unitary properties of the Fermi gas near
the resonance. The width of the phase slip is proportional to the
square root of the difference between the transition temperature
and the temperature. The signature of the phase slip in the
density profile becomes more robust across the BCS-BEC crossover.
\end{abstract}
\maketitle A lot of effort has been concentrated on creating the
superfluid state in a Fermi gas \cite{Pitaevskii}. In recent
years, the quantum degenerate regime was reached \cite{DeMarco}
and Feshbach resonance was observed in Fermi gases \cite{Loftus}.
Asymmetric expansions of cold Fermi gases near the resonance were
observed \cite{OHara, Regal2}, as a result of the strong
interaction between atoms \cite{Regal2}.  The molecular condensate
was observed when the scattering length was tuned to the positive
side of the resonance \cite{Jochim, Greiner}.  Condensation of
Fermion pairs was observed when the Fermi gas was fast tuned
through the resonance \cite{Regal, Zwierlein}.  The measurement of
the breathing-mode frequency was consistent with the predication
of superfluid hydrodynamics, and the damping of the mode was weak
at low temperatures \cite{Kinast}.  Across the Feshbach resonance,
a sudden change in the collective excitation spectrum was observed
\cite{Bartenstein}.  More recently, observation of the pairing gap
in the radio-frequency excitation spectra of the Fermi gas was
reported \cite{Chin}. However, phase coherence of the
wave-function, one of the most important properties of the
superfluid state, has not been observed in these experiments so
far. In this paper, we propose that phase coherence of the
superfluid state can be observed in the phase-slip experiment.

Phase coherence is one of the key characteristics of macroscopic
quantum phenomena such as Bose-Einstein condensation,
superfluidity, and superconductivity.  In these systems, the phase
symmetry of the wavefunction is broken.  When two systems are
coupled together, the two order parameters interfere with each
other.  The Josephson effect appears when two superconductors with
different phases are linked by a junction. In a system with two
Bose-Einstein condensates, the interference pattern was observed
upon release of the condensates\cite{andrews}.  In quasi-1D
systems, the order parameter in the middle region vanishes when
two sides of the system have different phases. This phenomenon is
called phase slip. In quasi-1D superconductors, phase slip is the
main reason for finite conductivity \cite{Lau}. In Bose-Einstein
condensates, phase slips have been observed in experiments
\cite{Burger, Denschlag}.  They are also often called solitons
because they have the same dynamics as solitons.

In the superfluid state of the Fermi gas, the phase slip can be
generated by the phase imprinting method.  A pulsed laser can
illuminate the system with different intensities on the left and
right sides, and the pulse duration can be tuned so that for an
atom the laser produces a phase difference of $\pi/2$ or $-\pi/2$
between the two sides. For a pair of atoms, the phase difference
is twice as much, $\pi$ or $-\pi$, between the two sides of the
system. As in superconductors, the order parameter in the
superfluid state describes the pairing of atoms. When there is a
discontinuity in the phase of a pair wave-function, the order
parameter must vanish at the point of discontinuity to generate a
phase slip.  Phase imprinting methods have been successfully
applied to create vortices \cite{Williams, Matthews} and solitons
\cite{Burger, Denschlag} in Bose-Einstein condensates.  However,
in the normal state of a Fermi gas, there is no phase coherence,
and the phase imprinting does not produce any special effect. In
the following, we study the properties of a phase slip in a
superfluid Fermi gas near a Feshbach resonance. It is assumed that
the phase step in the phase imprinting is applied along the axial
direction of the trap and the phase slip is generated at the
center of the trap.

In the superfluid state, the free energy of a Fermi gas is lower
than that in the normal state. This energy difference is defined
as the condensation energy.  The total free energy $L$ is the sum
of the condensation energy $L_C$ and the free energy of the normal
state $L_N$,
\begin{equation}
 L=L_C+L_N.
 \end{equation}
According to the Ginzburg-Landau theory, the condensation energy
can be expanded in terms of the order parameter $\phi$ which is
proportional to the energy gap of the fermions,
\begin{equation} \label{Lc}
L_C=\int d^3r \left[-{\hbar^2 \over 2m} \phi^*({\bf r}) \nabla^2
\phi({\bf r})+ \alpha({\bf r}) |\phi({\bf r})|^2+{\beta({\bf r})
\over 2} |\phi({\bf r})|^4 \right],
\end{equation}
where $\alpha$ and $\beta$ are coefficients.  Below the transition
temperature $T_C$, the sign of the coefficient $\alpha$ is
negative and the superfluid state has lower free energy than the
normal state.  The free energy of the normal state is generally a
function of temperature, particle density, trapping energy, and
interaction energy which is usually proportional to the scattering
length. The total free energy is a minimum in the space of the
particle density and the superfluid order parameter,
\begin{eqnarray}
{\delta L \over \delta n({\bf r})}=0, \label{s1} \\
{\delta L \over \delta \phi^{\ast}({\bf r})}=0, \label{s2}
\end{eqnarray}
where $n({\bf r})$ is the fermion particle density at location
${\bf r}$.  The derivative in equation (\ref{s1}) is taken under
the constraint of constant total-particle number. The equilibrium
configurations of the system can be determined from these
saddle-point equations.

In current experiments, the temperature is well below the Fermi
temperature $T_F$, and the system is often studied in the strong
interaction regime where the magnitude of the scattering length
can be much bigger than the interparticle distance. In this case,
the interparticle distance becomes the most important length
scale, and the thermodynamic properties of the Fermi gas are
unitary \cite{Ho}. To the leading order, the free energy density
of the normal state is given by
\begin{equation}
L_N=\int d^3r \left[ {\hbar^2 \over 2 m} c_N  n^{5 \over 3}({\bf
r})+V({\bf r}) n({\bf r}) \right ],
\end{equation}
where $V$ is the trapping potential and $c_N$ is a dimensionless
constant.  In the experiments \cite{OHara, Gehm}, about 25 percent
reduction of the energy by the interaction was observed,
corresponding to $c_N=4.3$.

In the following, we study the phase-slip phenomenon in the
unitary limit near the superfluid transition temperature, in the
region which is likely to be accessible in experiments.  The
coefficients $\alpha$ and $\beta$ in the unitary limit are
functions of the particle density $n$ and temperature $T$.  Near
the transition temperature, the coefficient $\alpha$ is
approximately given by
$$\alpha=c_{\alpha} k_B (T-T_C),$$ where $c_{\alpha}$ is a
dimensionless constant, $T$ is the temperature, and $k_B$ is the
Boltzmann constant.  In the unitary limit, the transition
temperature is given by $k_B T_C=c_T \hbar^2 n^{2/3}/(2 m)$, where
$c_T$ is a dimensionless constant. The theoretical estimation of
$T_C$ is about $0.5$ to $0.2$ times the Fermi temperature, and the
corresponding constant $c_T$ is between $4.8$ and $1.9$
\cite{Holland, Ohashi, Timmermans}. The coefficients $\beta$ in
equation (\ref{Lc}) is also a function of the fermion density,
$\beta=c_{\beta} \hbar^2/[2 m n^{1/3}]$, where $c_{\beta}$ is a
dimensionless constant.  Currently there are no theoretical
estimations of the constants $c_{\alpha}$ and $c_{\beta}$ for the
strong interaction in the unitary limit. However, the two
constants are available for the case of weak interactions
\cite{Popov},
\begin{eqnarray}
c_\alpha=1.47 c_T^2, \label{c_alpha}\\
c_\beta=2.94 c_T^2. \label{c_beta}
\end{eqnarray}
The renormalization of these two constants by the strong
interaction may not be significant, because the renormalization
reduces the Fermi energy only by about 25 percent \cite{OHara,
Gehm}.

In the unitary limit near the transition temperature, the saddle
point equations (\ref{s1}, \ref{s2}) are given by
\begin{eqnarray}
-{\hbar^2 \over 2 m} \nabla^2 \phi({\bf r})+\alpha({\bf r})
\phi({\bf r})+\beta({\bf r}) |\phi({\bf r})|^2
\phi({\bf r}) = 0, \label{OD} \\
{5 \hbar^2 \over 6 m} c_N n^{2 \over 3}({\bf r})+V({\bf
r})-\mu-{\hbar^2 \over 3 m} c_T c_\alpha {|\phi({\bf r})|^2 \over
n^{1 \over 3}({\bf r})}=0, \label{density}
\end{eqnarray}
where $\mu$ is the chemical potential.  In equation
(\ref{density}), the term quartic in the order parameter has been
dropped, because it is very small compared to other terms. In the
ground state, the phase of the order parameter is same everywhere.
Both the order parameter $\phi_0({\bf r})$ and the fermion density
$n_0({\bf r})$ of the ground state can be determined from these
equations.  The order parameter of the ground state varies slowly.
From equation (\ref{OD}), near the center of the trap, it is
approximately given by
\begin{equation}
\phi_0({\bf r}) \approx \sqrt{-{\alpha(0) \over \beta(0)}}=\sqrt{2
m c_\alpha k_B (T_C-T) n^{1/3}_0(0) \over c_\beta \hbar^2},
\end{equation}
where the origin is located at the center of the trap and $\phi_0$
is chosen to be positive.

In a phase-slip configuration, the phase of the order parameter
changes by $\pi$ from one side of the trap to the other side. With
this boundary condition, the order parameter $\phi_1({\bf r})$ and
the fermion density $n_1({\bf r})$ can also be obtained from the
saddle-point equations (\ref{OD}) and (\ref{density}).  Near the
transition temperature, the density of the order parameter
$|\phi|^2$ is much less than the fermion density. From equation
(\ref{density}), the difference between the fermion density in the
phase-slip configuration and that in ground state can be obtained,
\begin{equation} \label{dn}
\Delta n({\bf r}) \equiv n_1({\bf r})-n_0({\bf r}) \approx {3 c_T
c_\alpha \over 5 c_N} \Delta n_\phi({\bf r}),
\end{equation}
where $\Delta n_\phi$ is defined as $\Delta n_\phi({\bf r}) \equiv
|\phi_1({\bf r})|^2-|\phi_0({\bf r})|^2$.

\begin{figure}
\epsfxsize=12cm \centerline{ \epsffile{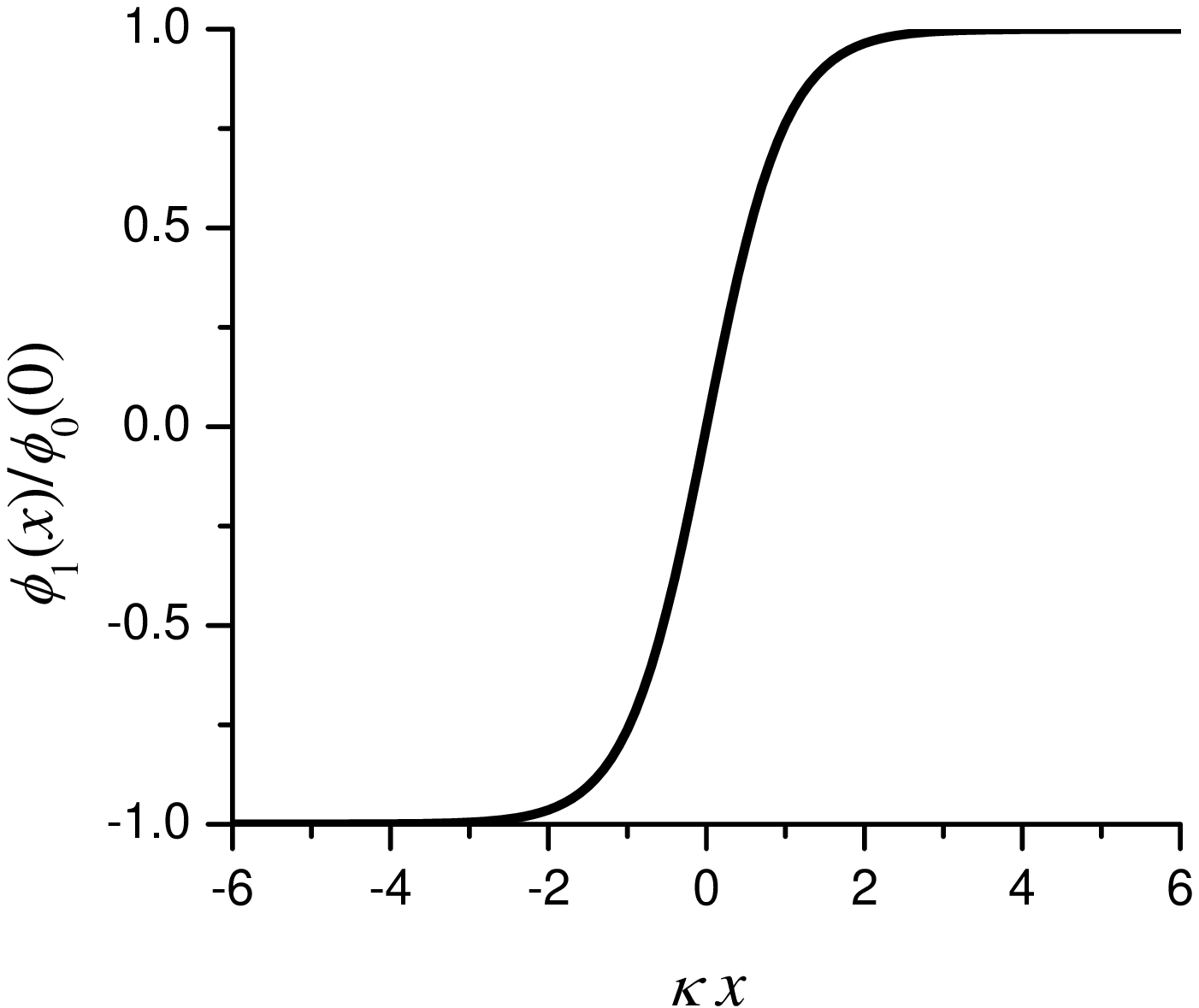} } \vspace{0.3cm}
\caption{\label{fig1}  The order parameter $\phi_1(x)$ of a
phase-slip configuration near the center of trap.  The width of
the phase slip is much smaller than the trap size.  The order
parameter vanishes in the middle. Across the middle, the phase
difference of the order parameter is given by $\pi$. Away from the
center, the magnitude of the order parameter approaches the value
of the ground state. }
\end{figure}

In a quasi-1D trap, the system is tightly trapped in two
directions, and the low-energy dynamics occurs in the other
direction. Even in a trap which is not quasi-1D such as the one in
Ref. \cite{Denschlag}, as long as the width of the phase slip is
much less than the trap size, the dynamics of the phase slip is
approximately one dimensional near the center of the trap. In this
case, equation (\ref{OD}) is reduced to a one-dimensional form,
\begin{equation}
-{\hbar^2 \over 2 m} {d^2 \phi(x) \over d x^2} +\alpha(x)
\phi(x)+\beta(x) |\phi(x)|^2 \phi(x) = 0,
\end{equation}
and the order parameter of the phase-slip state has an analytic
expression,
\begin{equation} \label{PS}
\phi_1(x) \approx \phi_0(0) \tanh (\kappa x),
\end{equation}
where $x$ is the coordinate along the axial direction, $\kappa
\equiv \sqrt{-m \alpha(0)} / \hbar$, and $\phi_1(x)$ is chosen to
be real.  As plotted in Fig. (\ref{fig1}), in a phase-slip
configuration, the order parameter vanishes in the middle.  There
is a phase difference of $\pi$ between the order parameter at one
side and that at the other side.  Far away from the middle, the
magnitude of the order parameter approaches the value of the
ground state. Note that the position of the phase slip does not
change with time, because the state given by equation (\ref{PS})
does not carry any current.

Near the center, according to equation (\ref{dn}), the density
change due to the phase slip is given by
\begin{equation}
\Delta n(x) \approx -{3 c_T c_\alpha \over 5 c_N} \phi_0^2(0)
\textrm{sech}^2 (\kappa x),
\end{equation}
and in terms of the density and temperature,
\begin{equation}
\Delta n(x) \approx {6 c_T c_\alpha^2 m\over 5 c_N c_\beta
\hbar^2} k_B (T-T_C) n^{1/3}_0(0) \textrm{sech}^2 (\kappa x).
\end{equation}
The magnitude of the density change $|\Delta n(x)|$ reaches the
maximum at the center and vanishes far away. The width of the
phase slip in the real space is of the order of $1/\kappa$, where
\begin{equation}
\kappa = \sqrt{ c_{\alpha} m k_B (T_C-T)} / \hbar.
\end{equation}
In experiments, if $\kappa$ and the temperature can be measured,
then the transition temperature $T_C$ can be determined. Right at
the center, the density change is proportional to the ratio of the
condensation-energy density ${\cal L}_C$ to the
normal-state-energy density ${\cal L}_N$,
\begin{equation}
{\Delta n(0) \over n_0(0)} \approx -{3 T_C {\cal L}_C(0) \over 5
(T-T_C) {\cal L}_N(0)},
\end{equation}
which is equal to
\begin{equation} \label{dratio}
{\Delta n(0) \over n_0(0)}={6 c_T c_\alpha^2 m k_B (T-T_C) \over 5
c_N c_\beta \hbar^2 n^{2/3}_0(0)},
\end{equation}
where ${\cal L}_C(0)=\alpha(0) |\phi_0(0)|^2/2$ and ${\cal
L}_N(0)=c_N n^{5/3}_0(0)$.  Thus the density profile of a phase
slip can provide important information about the properties of the
superfluid ground state such as the transition temperature and the
condensation energy.

In experiments, the method to detect the phase slip is the
measurement of the density profile. However, the trap also causes
the inhomogeneity of the density distribution. Both the phase slip
and the trap lead to the quadratic dependence of the density on
the distance away from the center, but they have different signs.
Whether or not the phase slip shows up as a dip in the density
profile is determined by the second derivative of the particle
density at the center.  The particle density can be written as the
sum of the particle density of the ground-state and the density
difference between the phase-slip state and the ground state,
\begin{equation}
n_1(x)=\Delta n(x)+n_0(x).
\end{equation}
The second derivative of the density change due to the phase
slip is given by
\begin{equation}
\Delta n''(0)={6 c_T c_\alpha \over 5 c_N} \phi_0^2(0)
\kappa^2={12 c_T c_\alpha^3 m^2 \over 5 c_N c_\beta \hbar^4} k_B^2
(T_C-T)^2 n_0^{1/3}(0).
\end{equation}
Near $T_C$, the density profile of the ground state is
approximately given by the Thomas-Fermi approximation,
\begin{equation}
{5 \hbar^2 \over 6 m} c_N n_0^{2 \over 3}(x)+V(x)-\mu=0,
\end{equation}
and the second derivative of the density is approximately given by
\begin{equation}
n_0''(0)=-{9m \over 5 \hbar^2 c_N} n_0^{1 \over 3}(0)
V''(0)=-{9m^2 \omega_x^2 \over 5 \hbar^2 c_N} n_0^{1 \over 3}(0),
\end{equation}
where $\omega_x$ is the axial frequency of the trap.  Whether the
phase slip appears as a dip in the density profile depends on the
ratio of the two second derivatives,
\begin{equation} \label{condition}
\left| {\Delta n''(0) \over n_0''(0)} \right|={4 c_T c_\alpha^3
k_B^2 (T-T_C)^2 \over 3 c_\beta \hbar^2 \omega_x^2}.
\end{equation}
If this ratio is bigger than one, then there is a dip in the
density profile.

\begin{figure}
\epsfxsize=12cm \centerline{ \epsffile{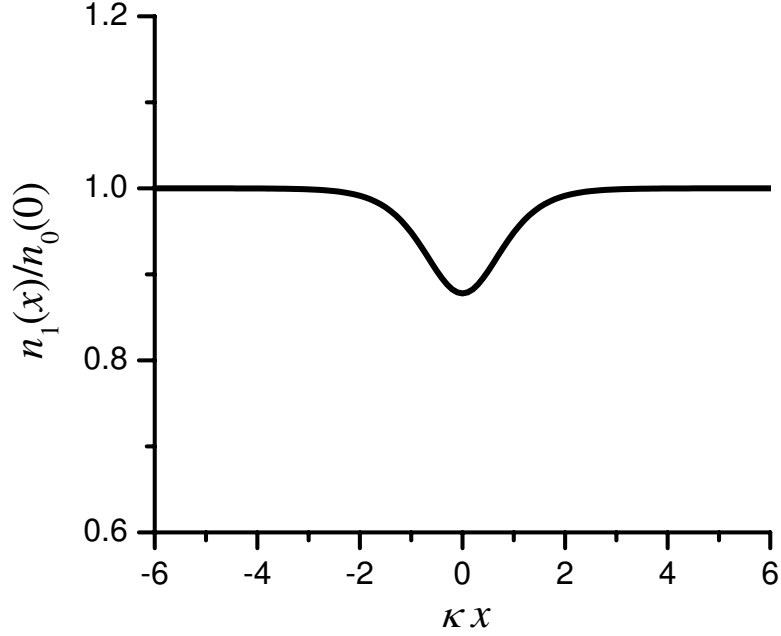} } \vspace{0.3cm}
\caption{\label{fig2}  The density profile of a phase slip for the
experimental conditions given in Ref. \cite{Chin} at the
temperature $T=0.9 T_C$, where $T_C \approx 0.25 T_F$,
$T_F=3.4\mu$K, and $\omega_x=24$Hz.  The units of the two axes are
given by $1/\kappa \approx 1/[0.39 n_0^{1 \over 3}(0)]$ and
$n_0(0) \approx 3.3 \times 10^{13}/cm^3$.  The value of $n_0(0)$
is calculated from the Thomas-Fermi approximation.  The width of
the phase slip is of the order of $1/\kappa$ which is much smaller
than the trap size.}
\end{figure}

In the experiment by Chin {\it et al.} \cite{Chin}, the excitation
gap was observed at the Fermi energy $E_F/k_B=3.2\mu$K with the
axial frequency $\omega_x/(2 \pi)=24$Hz.  The transition
temperature $T_C$ was estimated to be about a quarter of the Fermi
temperature \cite{Kinnunen}, corresponding to $c_T=2.4$. Currently
there are no better estimation of the constants $c_\alpha$ and
$c_\beta$ than the values given by equations (\ref{c_alpha}) and
(\ref{c_beta}). With these numbers, even if the temperature is
lower than the transition temperature by just one percent,
$T_C-T=0.01 T_C$, the ratio of the two density derivatives given
in equation (\ref{condition}) is already much larger than one,
$\Delta n''(0) / n_0''(0) \approx 6.8 \times 10^4$. Thus the phase
slip can be identified by the dip at the center of the trap in the
density profile for the vast range of the temperature below the
transition temperature.  In Fig. (\ref{fig2}), the density profile
of a phase slip at temperature $T=0.9 T_C$ is plotted near the
center of trap.  At the center, the particle density of the phase
slip is smaller than that of the ground state by about 12 percent,
which is very likely to be detected in the experiment. Although
better estimations of the constants such as $c_\alpha$ and
$c_\beta$ can improve these calculations, but it will probably not
lead to any change greater than an order of a magnitude, as the
interaction only changes the normal-state energy by about 25
percent in the unitary limit \cite{OHara, Gehm}.  The measurement
of the constants $c_\alpha$ and $c_\beta$ in the phase-slip
experiment can also serve as a test for microscopic theories.

The strong interaction is the reason for the unitary properties of
a Fermi gas.  However, when the Fermi gas is very close to a
Feshbach resonance, there is a crossover from the superfluid state
(BCS-type) to the molecular-condensate state (BEC).  In the
crossover region, fermion atoms coexist with diatomic molecules,
and properties of this mixture cannot be described by the unitary
properties of the fermions alone, because of the interaction
between atoms and molecules.  Although in the crossover region
many properties of the system are unclear, the phase slip can
still be generated by the phase-imprinting method, because in both
the superfluid and molecular-condensate states the phase symmetry
of the wavefunction is broken.

At the other side of the crossover, in the molecular-condensate
state with a weak interaction, the dynamics is well described by
the Gross-Pitaevski equation (GP-equation)
\begin{equation}
i\hbar {\partial \psi \over \partial t}=(-{\hbar^2 \over 4m}
\nabla^2+V) \psi+g_m |\psi|^2 \psi,
\end{equation}
where $\psi$ is the wavefunction of the molecular condensate,
$g_m$ is the coupling constant $g_m=\hbar^2 a_m/2m$, and $a_m$ is
the scattering length of the molecules. The soliton solution of
the GP-equation has been well studied.  Near the center of the
trap, the stationary soliton solution is approximately given by
\begin{equation}
\psi(x,t) \approx \psi_0 \tanh(\kappa_m x) e^{-2i \mu t/\hbar},
\end{equation}
where the chemical potential is given by $\mu=g_m |\psi_0|^2/2$
and $\kappa_m=2 \sqrt{m \mu}/\hbar$. In this configuration, the
molecular density vanishes at the center and the phase difference
of the wave-function across the center is given by $\pi$. The
width of the phase slip is of the order of $1/\kappa_m$ from which
both the coupling constant $g_m$ and the scattering length $a_m$
can be extracted. In Ref. \cite{Petrov}, $a_m$ was estimated to be
about 0.6 times the scattering length of the atoms.  The phase
slip in the molecular-condensate state is much easier to be
identified in the density profile because the density vanishes at
the center. Comparing the properties of the phase slip at both
sides of the BEC-BCS crossover, we expect that the particle
density at the center drops dramatically as the system is tuned
from the superfluid state to the molecular-condensate state, due
to the increase of the molecular density.

In conclusion, we predict that the phase slip can be generated in
a superfluid Fermi gas by the phase imprinting technique. The
presence of the phase slip can be detected by the measurement of
the density profile. The phase slip can provide an excellent
direct evidence of the superfluid state, because phase coherence
is one of the key signatures of superfluidity. The density profile
of a phase slip also offers important information about the
superfluid state, such as the condensation energy and the
transition temperature.  We would like to acknowledge support from
NSFC under Grant No. 90303008 and support from SRF for ROCS, SEM.

\newpage
\end{document}